\documentclass[english]{paper}
\usepackage[T1]{fontenc}
\usepackage[latin9]{inputenc}
\usepackage{units}
\usepackage{amsmath}
\usepackage{amssymb}
\usepackage{esint}

\makeatletter

\newcommand{\noun}[1]{\textsc{#1}}

\makeatother

\usepackage{babel}
\begin{document}

\title{Propagation of gravitational waves in a universe with slowly-changing
equation of state}

\author{Edmund Schluessel}

\date{July 24 2013}
\maketitle
\begin{abstract}
An exact solution for the expansion of a flat universe with dark energy
evolving according to a simple model is explored. The equation for
weak primordial gravitational waves propagating in this universe is
solved and explored; gravitational waves in a flat cosmology possessing
both a ``big bang'' singularity and a ``big rip'' singularity
can be described with confluent Heun functions. We develop approxmation
methods for confluent Heun equations in regimes of interest to gravitational
wave astronomers and predict the diminution in gravitational wave
amplitude in a universe with both a Big Bang and a Big Rip.
\end{abstract}

\section{Introduction}

\subsection{Background}

As we approach the beginnings of gravitational wave cosmology \cite{Probing spacetimes}
it becomes necessary to explore the full range of potential models
which might affect our initial observations. Riesse et al's \cite{Riees acceleration}
observation of acceleration greatly expanded the parameter space of
our models for cosmology but the paucity of observational data coupled
with the unknown character of the ``dark energy'' necessary to make
observations consistent with general relativity and the cosmological
principle have not helped narrow this space much in the past 16 years.
As the decay of short-wavelength gravitational waves is dependent
on the equation of state of the background through which they trave,
the amplitudes of these waves can be used as evidence for different
kinds of cosmic evolution. Some exploration has been done in this
field numerically \cite{Yang Zhang relic waves} but inconvenient
mathematics has made analytic work less common. Analytic work in this
field is necessary as a rapidly-evolving understanding of observational
data requires models which can be generalized and investigated for
easily characterizable phenomena. 

Throughout this paper we make the following assumptions:
\begin{itemize}
\item Classical, unmodified general relativity. We make no comment as to
the nature quantum-gravitational or other processes which might generate
primordial gravitational waves.
\item A cosmologically flat $K=0$ universe. While there is no absolute
evidence of the large-scale flatness of the universe, flatness is
a good approximation for closed models and remains consisten with
observations \cite{Planck}.
\item A universe which is, on large scales, homogeneous and isotropic.
\item The cosmological constant $\Lambda=0$. We will touch on cases where
the effective equation of state $w_{eff}=-1$, which are more general
than the cosmological constant-dominated de Sitter universe; as argued
in \cite{Babichev} and in \cite{Schluessel}, $w_{eff}=-1$ is a
limiting case for a number of cosmologies.
\end{itemize}

\subsection{Review: cosmologies with constant equation of state\label{sub:Review:-cosmologies-with}}

Consider a flat Friedmann-Lemaître-Robertson-Walker (FLRW) universe
described, for energy density $\epsilon$ and pressure \emph{p}, by
the Einstein equations%
\footnote{We use the convention $G=c=1$ throughout.%
}

\begin{equation}
G_{\mu\nu}=8\pi T_{\mu\nu}=8\pi\left(\begin{array}{cccc}
\epsilon & 0 & 0 & 0\\
0 & p & 0 & 0\\
0 & 0 & p & 0\\
0 & 0 & 0 & p
\end{array}\right).\label{eq:Einstein equations}
\end{equation}
Define the relationship $p/\epsilon\equiv w$, the ``equation of
state'' for a fluid filling the universe. We then have the well-known
Einstein equation relating the scale factor of the universe, the function
$a\left(t\right)$, to the equation of state%
\footnote{In this introductory section, we will use a dot to denote differentiation
with respect to time and a prime to denote differentiation with respect
to conformal time. In later sections we will use the comma derivative
to denote differentiation, so $f\left(x^{i}\right)_{,x^{j}}\equiv\partial f/\partial x^{j}$. %
}:

\begin{equation}
2\ddot{a}a+\left(1+3w\right)\dot{a}^{2}=0.\label{eq:master equation}
\end{equation}
In a flat homogemeous universe this relationship is true for any \emph{$w\left(t\right)$.
}Excluding the trivial solution $a=\mbox{constant}$, we arrive at
the general solution

\begin{equation}
a\left(t\right)=a_{0}e^{\frac{2}{3}\int\frac{1}{\int^{t}\left(1+w\right)d\bar{t}+C}dt}\label{eq:Scale factor eq}
\end{equation}
where $a_{0}$ is a positive number and \emph{C} an arbitrary constant.
Whether \emph{w} is constant or not, flat FLRW universes will have
Hubble parameter
\begin{equation}
\frac{\dot{a}}{a}\equiv H=\frac{2}{3}\frac{1}{\int\left(1+w\right)dt+C}\label{eq:Hubble parameter}
\end{equation}
and deceleration parameter
\begin{equation}
-\frac{\ddot{a}a}{\dot{a}^{2}}\equiv q=\frac{1+3w}{2}.\label{eq:deceleration parameter}
\end{equation}
The cases for a constant $w\equiv w_{0}>-1$ are well known and include
the ubiquitous ``matter-dominated'' ($w_{0}=0$) and ``radiation-dominated''
($w_{0}=1/3$) cosmologies thought, until the discovery of acceleration,
to provide a complete picture of the evolution of the post-inflationary
universe. The solutions for constant $w=w_{0}$ are given by

\begin{equation}
a\left(t\right)=\begin{cases}
a_{0}\left(t-t_{0}\right)^{\frac{2}{3\left(1+w_{0}\right)}} & w_{0}>-1\\
a_{0}e^{H_{0}t} & w_{0}=-1
\end{cases}\label{eq:-5}
\end{equation}
for some empirical constant Hubble parameter $H_{0}$ and where $t_{0}$,
another empirical constant, describes the time since a singularity
which for $w_{0}>-1$ represents a Big Bang.

The discovery by Riess et al. \cite{Riees acceleration} that $q<0$,
in sharp contrast to the prediction of $q=1/2$ for a universe with
$w=0$, has been confirmed multiple times \cite{Sako,Sollerman,Mikinatis,Sullivan,Sullivan 2,Drinkwater,Perlmutter,Kowalski,Wood-Vasey,Davis,Riess 2,Riess 3,Drinkwater 2},
implying the present day value of $w<-1/3$. A partial response to
this problem within the framework of FLRW cosmologies comes in the
form of the ``phantom energy''-dominated universe described in \cite{Phantom energy},
which has scale factor
\begin{equation}
a_{w<-1}\left(t\right)=a_{0}\left|t_{0}-t\right|^{\frac{2}{3\left(1+w_{0}\right)}}\label{eq:-6}
\end{equation}
for $w_{0}<-1$ and where the constant $t_{0}$, rather than denoting
a convergent singularity in the past, instead marks a divergent singularity
in the future -- a ``big rip'' where the scale factor goes to infinity
in finite time.

It is common to simplify the equations that arise in an expanding
universe by making a conformal change in variables as per \cite{Sachs-Wolfe}.
We define the new time variable $\eta$ by

\begin{equation}
dt\equiv ad\eta.\label{eq:}
\end{equation}
In conformal time, the equation describing the scale factor reads

\begin{equation}
a\left(\eta\right)=\begin{cases}
a_{0}\left|\eta-\eta_{0}\right|^{\frac{2}{1+3w_{0}}} & w_{0}\neq-1/3\\
a_{0}e^{H_{0}\eta} & w_{0}=-1/3
\end{cases}.\label{eq:ScaleFactorConformal}
\end{equation}
Note that as the two time coordinates are related by 
\begin{equation}
a_{0}\left(\eta-\eta_{0}\right)=\left(t-t_{0}\right)^{\frac{1+3w_{0}}{3\left(1+w_{0}\right)}}\label{eq:-10}
\end{equation}
the cases $-1/3>w_{0}>-1$ can be deceptive in terms of character
of the singularity when analysis is carried out in $\eta$-time. 

Indications are that the universe's acceleration is both recent and
increasing \cite{Suzuki}, that is, $\dot{q}<0$, implying the equation
of state is evolving with time. In \noun{section} \ref{sec:Evolution-of-a}
we will discuss a solution to \ref{eq:master equation} which obeys
this condition and which as a consequence has both ``big bang''
and ``big rip'' singularities.

\subsection{Gravitational waves in universes with constant equation of state\label{sub:Gravitational-waves-in}}

Equation (\ref{eq:Einstein equations}) admits a linear-order tensor
perturbation $\delta G_{\mu\nu}\left(\nu\left(t\right)\right)$ such
that

\begin{equation}
\ddot{\nu}+3\frac{\dot{a}}{a}\dot{\nu}+\frac{n^{2}}{a^{2}}\nu=0\label{eq:GW equation}
\end{equation}
\cite{Sachs-Wolfe,LL} where we have removed a gauge term by demanding
$\delta T_{\mu\nu}=0$. As $\nu$ is part of a tensorial solution
to the generalized Laplace equation, we regard $\nu$ as describing
a weak gravitational wave in the sense that $\left|\nu\right|\ll1$
such that $\nu^{2}\approx0$. In conformal time, (\ref{eq:GW equation})
becomes

\begin{equation}
\nu^{\prime\prime}+2\frac{a^{\prime}}{a}\nu^{\prime}+n^{2}\nu=0.\label{eq:GW conformal equation}
\end{equation}
When our equation of state $w=w_{0}$ this is expressed as
\begin{equation}
\begin{cases}
\nu^{\prime\prime}+\frac{4}{1+3w_{0}}\frac{1}{\eta_{*}}\nu^{\prime}+n^{2}\nu=0 & w_{0}\neq-1/3\\
\nu^{\prime\prime}+2H_{0}\nu^{\prime}+n^{2}\nu=0 & w_{0}=-1/3
\end{cases}\label{eq:-7}
\end{equation}
where we define $\eta_{*}=\left(\eta-\eta_{0}\right)\mbox{sgn}\left(1+w_{0}\right)$
and restrict our analysis only to the period of cosmological expansion
following (in the $w_{0}>-1$ case) or leading to (in the $w_{0}<-1$
case) the singularity. In the radiation-dominated $w_{0}=1/3$ universe,
this equation has the well-known solution
\begin{equation}
\nu_{\mbox{radiation}}=c_{1}\frac{\sin\left(n\eta_{*}\right)}{\eta_{*}}+c_{2}\frac{\cos\left(n\eta_{*}\right)}{\eta_{*}}=c_{1}j_{0}\left(n\eta_{*}\right)+c_{2}y_{0}\left(n\eta_{*}\right)\label{eq:-8}
\end{equation}
and in a matter-dominated $w_{0}=0$ universe \cite{Lifshitz}

\begin{eqnarray}
\nu_{\mbox{matter}} & = & \sqrt{\frac{2}{\pi n}}\left[c_{1}\left(\frac{\sin\left(n\eta_{*}\right)}{n\eta_{*}^{3}}-\frac{\cos\left(n\eta_{*}\right)}{\eta_{*}^{2}}\right)-c_{2}\left(\frac{\cos\left(n\eta_{*}\right)}{n\eta_{*}^{3}}+\frac{\sin\left(n\eta_{*}\right)}{\eta_{*}^{2}}\right)\right]\nonumber \\
 & = & \sqrt{\frac{2}{\pi n}}\frac{n}{\eta_{*}}\left[c_{1}j_{1}\left(n\eta_{*}\right)+c_{2}y_{1}\left(n\eta_{*}\right)\right].\label{eq:-9}
\end{eqnarray}
where $j_{\alpha}\left(x\right)$ and $y_{\alpha}\left(x\right)$
are spherical Bessel functions of the first and second kind \cite{A&S}.
The application of Bessel functions as a solution to (\ref{eq:GW conformal equation})
is trivial but seems not to be in wide discussion in the literature.
Therefore for completeness and as a reference to what follows we present
the solutions here with a few remarks.

For arbitrary constant $w$, (\ref{eq:GW conformal equation}) is
solved by 
\begin{eqnarray}
\nu_{w}\left(\eta_{*}\right) & = & \eta_{*}^{\frac{3w-3}{2+6w}}\left[c_{1}J_{\frac{1-3w}{1+3w}}\left(n\eta_{*}\right)+c_{2}Y_{\frac{1-3w}{1+3w}}\left(n\eta_{*}\right)\right]\nonumber \\
 & = & \sqrt{\frac{2n}{\pi}}\eta_{*}^{\frac{3w-1}{1+3w}}\left[c_{1}j_{\frac{1-3w}{1+3w}}\left(n\eta_{*}\right)+c_{2}y_{\frac{1-3w}{1+3w}}\left(n\eta_{*}\right)\right]\label{eq:Bessel solution}
\end{eqnarray}
except for $w=-1/3$ when (assuming $H_{0}^{2}\ll n^{2}$) 
\begin{equation}
\nu_{w=-1/3}\left(\eta_{*}\right)=e^{-H_{0}\eta}\left[c_{1}\sin\left(\sqrt{H_{0}^{2}-n^{2}}\eta_{*}\right)+c_{2}\cos\left(\sqrt{H_{0}^{2}-n^{2}}\eta_{*}\right)\right].\label{eq:Exponential solution}
\end{equation}
In the asymptotic limit of $n\eta_{*}\gg2\left|\left(1-3w_{0}\right)/\left(1+3w_{0}\right)^{2}\right|$,
(\ref{eq:Bessel solution}) and (\ref{eq:Exponential solution}) take
on the approximate form
\begin{equation}
v_{w_{0}}\left(\eta_{*}\right)\approx\frac{a_{0}}{a}\left[\bar{c}_{1}\sin\left(n\eta_{*}-\frac{\pi}{2}-\frac{\pi}{4}\frac{1-3w}{1+3w}\right)+\bar{c}_{2}\cos\left(n\eta_{*}-\frac{\pi}{2}-\frac{\pi}{4}\frac{1-3w}{1+3w}\right)\right],\label{eq:Approximate GW equation}
\end{equation}
in other words, the gravitational waves decay in amplitude proportional
to the scale factor, as broadly predicted in the theorem by Lifshitz
\cite{Lifshitz}. This result shows analytically the outcome predicted
numerically by \cite{Yang Zhang relic waves}, in particular that
a gravitational wave will decay less in a phantom energy-driven universe
than in a universe expanding through the influence of ordinary matter.
The Bessel functions have the property of a regular singularity at
$\eta_{*}=0$ and an irregular singular point at $\eta_{*}=\infty$.

By re-stating (\ref{eq:Approximate GW equation}) in \emph{t}-time,
it is also obvious gravitational waves in a universe with constant
equation of state undergo the same redshift as electromagnetic waves.
Noting frequency \emph{$f\left(t\right)\propto a^{-1}$},
\begin{equation}
\frac{f_{\mbox{emission}}}{f_{\mbox{observation}}}=\frac{a_{\mbox{observation}}}{a_{\mbox{emission}}}.\label{eq:-1}
\end{equation}

The amplitude \emph{A }of the waves meanwhile diminishes in proportion
to the scale factor, 

\begin{equation}
\frac{A_{\mbox{observation}}}{A_{\mbox{emission}}}=\frac{a_{\mbox{observation}}}{a_{\mbox{emission}}}.\label{eq:-1-1}
\end{equation}

\section{Evolution of a universe with simply-changing equation of state\label{sec:Evolution-of-a}}

\subsection{Cosmologies for a linear-function equation of state}

\subsubsection{Solutions in \textmd{\emph{t}}\textmd{-time}}

Cosmological evolution for a flat universe with constant \emph{w}
is well-known. We will consider the evolution of a universe with a
slightly more complicated equation of state: 
\begin{equation}
w=w_{0}-w_{1}t;\label{eq:EOS t time}
\end{equation}
and in conformal time, the related universe with 
\begin{equation}
w=w_{0}-v_{1}\eta.\label{eq:EOS eta-time}
\end{equation}
where $w_{1}$ and $v_{1}$ are arbitrary real constants (we do \emph{not}
assume that $w_{1}$ and $v_{1}$ are necessarily small). The solutions
for cosmologies with the above equations of state have been explored
by Babichev \emph{et al}~\cite{Babichev} in the context of purely
dark energy-dominated cosmologies; we replicate their work in a more
practical formalism in this section and apply it in \noun{subsection}
\ref{sub:Can-the-double-singularity} and \noun{section} \ref{sec:Gravitational-waves-in}.

Substituting (\ref{eq:EOS t time}) into (\ref{eq:Scale factor eq})
and solving gives us the generic solution
\begin{equation}
a\left(t\right)=\left[e^{\int\frac{1}{-\frac{1}{2}w_{1}t^{2}+\left(1+w_{0}\right)t+C}dt}\right]^{2/3}\label{eq:-2}
\end{equation}
 but this must be analyzed in cases to get useful results. These cases
are distinguished by the discriminant $\Delta\equiv\left(1+w_{0}\right)^{2}+2w_{1}C$
of the integral $\int wdt$:

\begin{equation}
a=\begin{cases}
a_{0}e^{\frac{4}{3\sqrt{-\Delta}}\tan^{-1}\left(\frac{-w_{1}t+\left(1+w_{0}\right)}{\sqrt{2w_{1}C-\left(1+w_{0}\right)^{2}}}\right)} & \Delta<0\\
a_{0}e^{-\frac{4}{3}\frac{1}{-w_{1}t+\left(1+w_{0}\right)}} & \Delta=0\\
a_{0}\left(-\frac{2}{w_{1}}\frac{-w_{1}t+\left(1+w_{0}\right)-\sqrt{\Delta}}{-w_{1}t+\left(1+w_{0}\right)+\sqrt{\Delta}}\right)^{\frac{2}{3}\frac{1}{\sqrt{\left(1+w_{0}\right)^{2}-2w_{1}C}}} & \Delta>0.
\end{cases}\label{eq:Cosmologies}
\end{equation}
Each of these, for appropriate choices of $w_{0}$, $w_{1}$ and \emph{C},
produces a plausible universe, that is, one with an epoch where \emph{H}
is positive.

Cosmologies of the first kind in (\ref{eq:Cosmologies}) have no singularities
in their evolution, but evolve from an initial finite non-zero scale
factor in the infinite past to another in the infinite future, the
two scale factors separated by a multiplicative factor of $\exp\left[4\pi/\left(3\sqrt{-2w_{1}C-\left(1+w_{0}\right)^{2}}\right)\right]$.
The parameter \emph{C}~cannot be set to zero in such a universe and
controls the ratio between the maximum and minimum scale factor. Cosmologies
of this kind model an expanding universe only when $w_{1}>0$.

Cosmologies of the second kind in (\ref{eq:Cosmologies}) have the
tuned value $C=\left(1+w_{0}\right)^{2}/2w_{1}$. This solution to
(\ref{eq:master equation}), when and only when $w_{1}<0$, has two
regions which could model an expanding universe. One expands from
an initial finite non-zero scale factor in the infinite past to a
big rip at $t=\left(1+w_{0}\right)/w_{1}$; the other expands asymptotically
from a big bang at $t=\left(1+w_{0}\right)/w_{1}$ to a finite scale
factor in the infinite future.

A cosmology of the third kind in (\ref{eq:Cosmologies}) can evolve
in a manner most like that of our own universe. In this model the
parameter \emph{C} can be set to zero without loss of generality,
giving a simplified expression of
\begin{equation}
a=a_{0}\left(\frac{2t}{2\left(1+w_{0}\right)-w_{1}t}\right)^{\frac{2}{3}\frac{1}{1+w_{0}}}.\label{eq:simplified two-singularity cosmology}
\end{equation}
When $w_{1}>0$ this cosmology has a region marked by an initial convergent
``big bang'' singularity, a period of expansion, and a divergent
``big rip'' singularity; where $w_{0}>-1$ the ``big bang'' takes
place at $t=0$. When $w_{1}<0$ this cosmology contains two different
regions of expansion similar to that of the cosmology of the second
kind.

\subsubsection{Related models in conformal time}

While the relationship between coordinates \emph{t} and $\eta$ is
generally not simple, we can work out solutions to (\ref{eq:master equation})
analogous to (\ref{eq:Cosmologies}) by use of the model equation
of state (\ref{eq:EOS eta-time}). These solutions are simply, where
$\bar{\Delta}\equiv\left(1+3w_{0}\right)^{2}+6v_{1}\bar{C}$

\begin{equation}
a\left(\eta\right)=\begin{cases}
a_{0}e^{\frac{4}{\sqrt{-\bar{\Delta}}}\tan^{-1}\left(\frac{1+3w_{0}-3v_{1}\eta}{\sqrt{-\bar{\Delta}}}\right)} & \bar{\Delta}<0\\
a_{0}e^{\frac{4}{3v_{1}\eta-\left(1+3w_{0}\right)}} & \bar{\Delta}=0\\
a_{0}\left(\frac{2\eta}{2\left(1+3w_{0}\right)-3v_{1}\eta}\right)^{\frac{2}{1+3w_{0}}} & \bar{\Delta}>0
\end{cases}\label{eq:eta cosmologies}
\end{equation}
where in the third case we have set $\bar{C}=0$ analogously to case
(\ref{eq:simplified two-singularity cosmology}) above.

\subsection{Can the double-singularity model describe our universe?\label{sub:Can-the-double-singularity}}

While the cosmology of the third kind in (\ref{eq:Cosmologies}) is
qualitatively desirable to model a universe with slowly-evolving dark
energy, it should also be quantitatively compatible with observations.
For an equation of state (\ref{eq:EOS t time}) and setting $w_{0}=0$,
the system created by (\ref{eq:Hubble parameter}) and (\ref{eq:deceleration parameter})
is solved by
\begin{eqnarray}
w_{1} & = & -\frac{H_{0}}{6}\left(2q_{0}-1\right)\left(q_{0}+\frac{5}{2}\right)\label{eq:-3}\\
t_{\mbox{now}} & = & \frac{2}{H_{0}\left(q_{0}+\frac{5}{2}\right)}.\label{eq:-4}
\end{eqnarray}
For the current best values for the Hubble parameter $H_{0}=\unitfrac[67.80\pm0.77]{km}{s\times mPc}$
\cite{H0} and $q_{0}=-0.53_{-0.13}^{+0.17}$ \cite{q0} we obtain

\begin{eqnarray}
w_{1} & = & \unit[1.5_{-0.0}^{+0.1}\times10^{-18}]{s^{-1}}\nonumber \\
t_{\mbox{now}} & = & \unit[4.6\pm0.4\times10^{17}]{seconds}=\unit[15\pm1]{Gya}\label{eq:age estimate}\\
t_{\mbox{rip}} & = & \unit[1.3_{-0.0}^{+0.1}\times10^{18}]{seconds}=\unit[41\pm3]{Gya}\nonumber 
\end{eqnarray}
which is easily compatible with the current observations for the age
of the universe \cite{H0} and indicates roughly 37\% of the universe's
lifespan has passed before a ``big rip'' 26 billion years in the
future.

The $\eta$-time model is more limited than our \emph{t}-time model:
the central, ``bang-rip'' region of the solution can only explain
$-1/2<q<1/2$ if we assume $w_{0}=0$. Nonetheless it will serve our
purpose in beginning the exploration of gravitational waves in such
cosmologies.

\section{Gravitational waves in a universe with simply-changing equation of
state\label{sec:Gravitational-waves-in}}

\subsection{Evolution of gravitational waves in $\eta$-time}

The origin of the universe, whatever form it took, likely left an
imprint in the form of a gravitational wave background generated by
the universe's primordial physical processes. The upcoming Next Gravitational-wave
Observatory (NGO, formerly LISA) should be able to detect high-frequency
relic gravitational waves \cite{eLisa}. We have seen in \noun{subsection}
\ref{sub:Gravitational-waves-in} how a universe with constant equation
of state preserves the spectrum of relic gravitational waves over
time. The same will be true in a universe with an evolving equation
of state.

Plugging the third case of (\ref{eq:eta cosmologies}) into (\ref{eq:GW equation})
gives us

\begin{equation}
\nu_{,\eta,\eta}+\frac{4}{1+3w_{0}}\frac{1}{\eta\left(1-\frac{3v_{1}}{2\left(1+3w_{0}\right)}\eta\right)}\nu_{,\eta}+n^{2}\nu=0.\label{eq:eta wave}
\end{equation}
With a changes of variables

\begin{equation}
\xi\equiv\frac{3v_{1}}{2\left(1+3w_{0}\right)}\eta\label{eq:-11}
\end{equation}
we obtain
\begin{equation}
\xi\left(\xi-1\right)\nu_{,\xi,\xi}-2\left(1+\beta\right)\nu_{,\xi}+\omega^{2}\xi\left(\xi-1\right)\nu=0\label{eq:xi form}
\end{equation}
where 
\begin{eqnarray}
\beta & \equiv & \frac{1-3w_{0}}{1+3w_{0}}\label{eq:-16}\\
\omega & \equiv & 2\left(1+3w_{0}\right)n/3v_{1}.\label{eq:-14}
\end{eqnarray}
We recognize (\ref{eq:xi form}) as being in the form of the Generalized
Spherical Wave Equation (GSWE) \cite{Leaver}. If we go on to change
the independent variable 

\begin{equation}
\mu=e^{-\imath\omega\xi}\nu\label{eq:-15}
\end{equation}

we obtain, where $\imath$ is the imaginary unit, 
\begin{equation}
\mu_{,\xi,\xi}+\left(2\imath\omega+\frac{2\left(1+\beta\right)}{\xi}-\frac{2\left(1+\beta\right)}{\xi-1}\right)\mu_{,\xi}-\frac{2\imath\left(1+\beta\right)\omega}{\xi\left(\xi-1\right)}\mu=0\label{eq:Heun wave}
\end{equation}
which we recognize as the confluent Heun equation \cite{Slavyanov}.

Formally, we can say that (\ref{eq:Heun wave}) has particular solution
\begin{equation}
\mu_{1}=\mbox{Hc}\left(\begin{array}{ccc}
\frac{1}{2}\imath\omega, & 2\left(1+\beta\right), & -2\left(1+\beta\right)\\
0, & 2\imath\left(1+\beta\right)\omega
\end{array};\xi\right)\label{eq:Heun solution}
\end{equation}
and therefore (\ref{eq:eta wave}) has general solution
\begin{equation}
\nu\left(\eta\right)=\left\{ \begin{array}{c}
e^{\imath n\eta}\mbox{Hc}\left(\begin{array}{ccc}
\frac{1}{2}\imath\omega, & 2\left(1+\beta\right), & -2\left(1+\beta\right)\\
0, & 2\imath\left(1+\beta\right)\omega
\end{array};\frac{3v_{1}}{2\left(1+3w_{0}\right)}\eta\right)\times\\
\times\left[C_{1}+C_{2}\int^{\eta}\frac{e^{-2\imath n\chi}a^{2}\left(\chi\right)}{\left[\mbox{Hc}\left(\begin{array}{ccc}
\frac{1}{2}\imath\omega, & 2\left(1+\beta\right), & -2\left(1+\beta\right),\\
0, & 2\imath\left(1+\beta\right)\omega
\end{array};\frac{3v_{1}}{2\left(1+3w_{0}\right)}\chi\right)\right]^{2}}d\chi\right]
\end{array}\right\} \label{eq:-13}
\end{equation}
where Hc is the confluent Heun function $\mbox{Hc}\left(p,\gamma,\delta,\alpha,\sigma;z\right)$
as defined in \cite{Slavyanov}. This solution is not especially useful
in and of itself, as (\ref{eq:Heun solution}) cannot generally be
expressed as a finite series of simpler functions \cite{Slavyanov,Leaver}
although much work has been put into infinite-series expressions for
Heun functions (see for example \cite{Confluent solutions,Shahnazaryan,Fiziev}).
Moreover as the most physically interesting cases of $\beta$ are
integer values $\beta=\left\{ 0,1\right\} $ corresponding to $w_{0}=\left\{ 1/3,0\right\} $,
we need to use the Frobenius method to express the second solution.

Leaver \cite{Leaver} provides the oldest series solutions for $\mu$
about the points $\xi=0$ and $\xi=1$, those of Baber and Hassé.
Equation (\ref{eq:Heun wave}) has particular three-term recurrent
series solution
\begin{eqnarray}
\mu_{1} & = & \sum_{m=0}^{\infty}b_{m}\xi^{m}\label{eq:-18}\\
b_{1}+\imath\omega b_{0} & = & 0\\
\left\{ \begin{array}{c}
-\left(m+2\right)\left(m+3+2\beta\right)b_{m+2}+\\
+\left[m^{2}+\left(1-2\imath\omega\right)m-2\left(2+\beta\right)\imath\omega\right]b_{m+1}+\\
+2\imath\omega mb_{m}
\end{array}\right\}  & = & 0\label{eq:-17}
\end{eqnarray}
about $\xi=0$. The ``decaying'' mode of the gravitational waves
in our model universe diverges as $\eta^{-1-2\beta}$. In contrast
to the solutions for a universe with constant equation of state, the
second solution includes a $\left(\ln\eta\right)$-term when $2\beta$
is an integer; this solution can be obtained by the Frobenius method.

While in general (\ref{eq:xi form}) is difficult to work with analytically,
in the particular case of high-frequency gravitational waves, that
is $\omega\gg1$, we can say to good approximation
\begin{equation}
\nu\approx\left(\xi-1\right)^{1+\beta}\left(\omega\xi\right)^{-\beta}\left[C_{1}j_{\beta}\left(\omega\xi\right)+C_{2}y_{\beta}\left(\omega\xi\right)\right];\label{eq:Heun approximation}
\end{equation}
for some constants $C_{1},C_{2}$; this approximation is fully derived
in the \noun{appendix}. This approximation (\ref{eq:Heun approximation})
is nearly identical in form to (\ref{eq:Bessel solution}); indeed,
recalling $\omega\xi=n\eta$ we can go on to approximate
\begin{equation}
\nu\approx a^{-1}\left[C_{1}\sin\left(n\eta-\frac{\pi}{2}\left(1+\beta\right)\right)+C_{2}\cos\left(n\eta-\frac{\pi}{2}\left(1+\beta\right)\right)\right].\label{eq:Heun final eta approximation}
\end{equation}

The solution for $\nu$, at least in approximation, goes to zero in
finite time. This implies that gravitational waves in a universe expanding
with scale factor as in the third case of (\ref{eq:eta cosmologies})
must always decay faster than those with identical initial conditions
in a universe with analogous constant equation of state. Because the
time parameter $\eta$ is defined by this scale factor function, though,
we cannot easily compare the functions directly and need to return
to \emph{t}-time in order to best discuss how our dark energy model
would affect gravitational wave observations.

\subsection{Evolution of gravitational waves in \emph{t}-time\label{sub:Evolution-of-gravitational}}

In universes with a single singularity as discussed in \noun{subsection}
\ref{sub:Review:-cosmologies-with}, relating solutions to the wave
equation (\ref{eq:GW equation}) to corresponding solutions in conformal
time is trivial, making solving the equations in $\eta$ mathematically
desirable. In the cosmologies under our consideration, the relationship
between \emph{t} and $\eta$ is more complicated, requiring the use
of inverse functions not expressible in terms of standard functions
and producing no mathematical advantages in the solution of the associated
differential equations. Therefore we briefly explore solutions to
(\ref{eq:GW equation}) in \emph{t-}time for our equation of state
(\ref{eq:EOS t time}).

In \emph{t}-time, our gravitational wave equation for a a universe
with equation of state (\ref{eq:EOS t time}) reads
\begin{equation}
\ddot{\nu}+\frac{4}{t\left[2\left(1+w_{0}\right)-w_{1}t\right]}\dot{\nu}+\frac{n^{2}}{a_{0}^{2}}\left(\frac{2t}{2\left(1+w_{0}\right)-w_{1}t}\right)^{-\frac{4}{3}\frac{1}{1+w_{0}}}\nu=0.\label{eq:-19}
\end{equation}
Introducing the following notation to simplify our expressions:
\begin{eqnarray}
x & \equiv & \frac{w_{1}}{2\left(1+w_{0}\right)}t\label{eq:-27}\\
\upsilon & \equiv & \left(1+w_{0}\right)\frac{n}{a_{0}}\left(\frac{w_{1}}{2}\right)^{\frac{3+2\beta}{2+\beta}}\label{eq:-28}
\end{eqnarray}
(the scale factor takes on the simple form 
\begin{equation}
a=a_{0}\left(\frac{2}{w_{1}}\frac{x}{1-x}\right)^{\frac{1+\beta}{2+\beta}}\label{eq:-33}
\end{equation}
in this notation) we arrive at
\begin{equation}
x\left(x-1\right)\nu_{,x,x}-3\frac{1+\beta}{2+\beta}\nu_{,x}-\upsilon^{2}\left(1-x\right)^{\frac{4+3\beta}{2+\beta}}x^{-\frac{\beta}{2+\beta}}\nu=0.\label{eq:delta x}
\end{equation}
In the case of $w_{0}=1/3\iff\beta=0$, this gives us another GSWE:
\begin{equation}
x\left(x-1\right)\nu_{,x,x}-\frac{3}{2}\nu_{,x}-\upsilon^{2}\left(1-x\right)^{2}\nu=0\label{eq:rad heun t}
\end{equation}
with corresponding solution
\begin{equation}
\nu=e^{\upsilon x}\left[\begin{array}{c}
C_{1}\mbox{Hc}\left(\begin{array}{ccc}
\upsilon/2, & 3/2, & -3/2,\\
\upsilon/2, & 3\upsilon/2+\upsilon^{2},
\end{array};x\right)+\\
+C_{2}\left(x-1\right)^{5/2}\mbox{Hc}\left(\begin{array}{ccc}
\upsilon/2, & 3/2, & 7/2,\\
5/2+\upsilon/2, & \upsilon^{2}+3\upsilon/2+15/4,
\end{array};x\right)
\end{array}\right]\label{eq:rad heun}
\end{equation}
(note the solution is separated into a confluent Heun function with
purely real parameters and an exponential function with purely real
power, in contrast to the analogous $\eta$-time case). This solution
can be approximated, using the same method as for (\ref{eq:Heun approximation}),
as
\begin{equation}
\nu\approx\left(\frac{1-x}{x}\right)^{\frac{1}{2}}\left[\begin{array}{c}
c_{1}\cos\left(\frac{1}{4}\pi-\frac{1}{2}\pi\upsilon+\upsilon\sqrt{x\left(1-x\right)}+\upsilon\sin^{-1}\sqrt{x}-\frac{1}{\upsilon}\frac{\sqrt{x}\left(x-6\right)}{48\left(1-x\right)^{3/2}}\right)+\\
+c_{2}\cos\left(\frac{1}{4}\pi-\frac{1}{2}\pi\upsilon-\upsilon\sqrt{x\left(1-x\right)}-\upsilon\sin^{-1}\sqrt{x}+\frac{1}{\upsilon}\frac{\sqrt{x}\left(x-6\right)}{48\left(1-x\right)^{3/2}}\right)
\end{array}\right].\label{eq:rad heun approx}
\end{equation}

Also of note is the phantom-energy case $w_{0}=-7/3\iff\beta=-4/3$,
which gives GSWE
\begin{equation}
x\left(x-1\right)\nu_{,x,x}+\frac{3}{2}\nu_{,x}-\upsilon^{2}x^{2}\nu=0\label{eq:-21}
\end{equation}
and therefore has exact solution

\begin{equation}
\nu=e^{\upsilon x}\left[\begin{array}{c}
C_{1}\mbox{Hc}\left(\begin{array}{ccc}
\upsilon/2, & -3/2, & 3/2,\\
-\upsilon/2, & -3\upsilon/2,
\end{array};x\right)+\\
+C_{2}\left(x-1\right)^{-1/2}\mbox{Hc}\left(\begin{array}{ccc}
\upsilon/2, & -3/2, & 1/2,\\
-\left(1+\upsilon\right)/2, & 3\upsilon/2+3/4,
\end{array};x\right)
\end{array}\right].\label{eq:phantom heun}
\end{equation}

The case of greatest physical interest, $w_{0}=0\iff\beta=1$, does
not give us an equation apparently solvable in terms of Heun functions
but we can write the approximate solution 
\begin{equation}
\nu\approx\left(\frac{1-x}{x}\right)^{\frac{1}{2}\frac{1+2\beta}{2+\beta}}\left[\begin{array}{c}
C_{1}J_{\beta+\frac{1}{2}}\left(\left(2+\beta\right)\upsilon\left(\frac{x}{1-x}\right)^{\frac{1}{2+\beta}}\right)+\\
+C_{2}Y_{\beta+\frac{1}{2}}\left(\left(2+\beta\right)\upsilon\left(\frac{x}{1-x}\right)^{\frac{1}{2+\beta}}\right)
\end{array}\right]\label{eq:General t solution}
\end{equation}
which can be further approximated as 
\begin{equation}
\nu\approx a^{-1}\left[c_{1}\cos\left(\left(2+\beta\right)\upsilon\left(\frac{x}{1-x}\right)^{\frac{1}{2+\beta}}+\frac{\pi}{2}\beta\right)+c_{2}\sin\left(\left(2+\beta\right)\upsilon\left(\frac{x}{1-x}\right)^{\frac{1}{2+\beta}}+\frac{\pi}{2}\beta\right)\right]\label{eq:General t approximation}
\end{equation}
and in the case of $w_{0}=0\iff\beta=1$ we have
\begin{equation}
\nu\approx\left(\frac{1-x}{x}\right)^{1/2}\left[C_{1}J_{3/2}\left(3\upsilon\left(\frac{x}{1-x}\right)^{1/3}\right)+C_{2}Y_{3/2}\left(3\upsilon\left(\frac{x}{1-x}\right)^{1/3}\right)\right].\label{eq:w=00003D0 solution}
\end{equation}
This approximation suggests the frequency of gravitational waves will
decrease with time until $x=2/3$ after which time the frequency diverges
to infinity as the amplitude converges to zero. This represents an
artefact of the approximation scheme rather than a gravitational ``ultraviolet
catastrophe'', as will be seen in \noun{subsection }\ref{sub:Gravitational-waves-at}.

\subsection{Amplitude of gravitational waves}

From (\ref{eq:age estimate}) and (\ref{eq:General t solution}-\ref{eq:w=00003D0 solution})
we are prepared to make quantitative predictions about the evolution
of gravitational waves in our accelerating universe. Let $A=A\left(\nu\left(t^{*}\right)\right)$
denote the RMS amplitude of a gravitational wave described by $\nu\left(t\right)$
over one cycle evaluated in an interval about some time $t^{*}$;
let $F=F\left(\nu\left(t^{*}\right)\right)$ denote the reciprocal
of the period of a cycle of the wave evaluated on this same interval.
Let symbols with a bar denote evaluations made in an accelerating
universe modeled by (\ref{eq:simplified two-singularity cosmology})
and let symbols with no bar denote evaluations made in a corresponding
Friedmann universe with identical parameters except $w_{1}=0$. Then
the following relation describes the decay of gravitational waves
in our universe versus that in a classical Friedmann universe:

\begin{eqnarray}
\frac{\bar{A}_{\mbox{observed}}/\bar{A}_{\mbox{emitted}}}{A_{\mbox{observed}}/A_{\mbox{emitted}}} & = & \left(\frac{1-x_{\mbox{observed}}}{1-x_{\mbox{emitted}}}\right)^{\frac{1+\beta}{2+\beta}}\label{eq:-25}\\
 & = & \left[\frac{2\left(1+w_{0}\right)-w_{1}t_{\mbox{observed}}}{2\left(1+w_{0}\right)-w_{1}t_{\mbox{emitted}}}\right]^{\frac{2}{3}\frac{1}{1+w_{0}}}.\label{eq:-12}
\end{eqnarray}

By (\ref{eq:age estimate}) we live at $x\approx.37$, so a primordial
gravitational wave will have a decayed to a strength, relative to
that in the analogous FLRW case, of
\begin{equation}
\bar{A}/A\approx\left(1-w_{1}t/2\right)^{2/3}\sim.73,\label{eq:-23}
\end{equation}
in other words, a gravitational wave created at $t_{\mbox{emitted}}\approx0$
and evolving in our accelerating universe will have only 73\% the
amplitude today it would have in a universe evolving without dark
energy.

\subsection{Gravitational waves at the ``big rip'' singularity\label{sub:Gravitational-waves-at}}

As we inhabit a universe closer to its beginning than its end, for
purposes of astronomy we were, in the previous subsection, primarily
interested in approximations for approximations about $x=0$. In terms
of mathematical techniques, though, the two regular singularities
of a Heun function at $x=0$ and $x=1$ are identical and therefore
we can just as easily discuss the fate of gravitational waves as the
scale factor diverges to infinity.

For sake of simplicity consider the case $w_{0}=1/3\iff\beta=1$.
Applying the same techniques as brought us from (\ref{eq:rad heun t})
to (\ref{eq:rad heun approx}) approximating about $x=1$ we have
\begin{equation}
\nu\approx e^{\upsilon\left(1-x\right)}x^{3/4}\left\{ \begin{array}{c}
C_{1}\left[\phantom{}_{1}F_{1}\left(-\frac{\upsilon}{2}-\frac{3}{4};-\frac{3}{2};2\upsilon\left(x-1\right)\right)\right]+\\
+C_{2}\psi\left(-\frac{\upsilon}{2}-\frac{3}{4};-\frac{3}{2};2\upsilon\left(x-1\right)\right)
\end{array}\right\} .\label{eq:-24}
\end{equation}
where $\phantom{}_{1}F_{1}\left(a,b,w\right)$ is Kummer's degenerate
hypergeometric function and $\psi\left(a,b,w\right)$ is Tricomi's
function \cite[chapter 13]{A&S}. Both branches of this function are
continuous in their derivatives at $x=1$. This result is at first
glance surprising, as it implies that gravitational waves at the Big
Rip diminish to zero amplitude and zero frequency -- but then resurge,
increasing in frequency and amplitude and carrying information across
the singularity and into whatever comes afterward!

Such a result is not so alarming in the wider context of gravitational
physics however. The extensibility of a metric across a regular singularity
is the motivation behind the well-known Kruskal representation of
the Schwarzschild solution describing a black hole \cite{MTW}. The
analogy between the Big Rip and black hole physics is qualitatively
obvious -- what, after all, is the Big Rip other than an infinite-redshift
surface?

\section{Conclusions}

Our statement of the scale factor of a universe driven by slowly-evolving
dark energy is mathematically more convenient than the form given
in \cite{Babichev}. The two-singularity Big Bang-Big Rip model is
more realistic than the Big Rip cosmology in its purest, $w=\mbox{constant}$
form. The current data also suggests the Bang-Rip cosmology is equally
plausible as one with only a Big Rip.

It is frequently the case that developments in physics and mathematics
run neck-and-neck. The beginning of the recent more widespread exploration
of Heun functions (1995, with \cite{Ronveaux}) coincides more or
less with the discovery of cosmic acceleration \& dark energy (1998,
with \cite{Riees acceleration}) so from a philosophical standpoint
our result in \noun{section }\ref{sec:Gravitational-waves-in} is
not surprising. Greater awareness of this class of functions will
lead to greater exploration within the community of mathematical physicists
of systems with multiple singularities.

While the diminution in gravitational wave amplitude we predict in
\noun{subsection \ref{sub:Evolution-of-gravitational} }is not great,
it should be detectable by planned gravitational wave observatories.
As predicted by \cite{Probing spacetimes}, gravitational waves can
be used as standard sirens for measuring the cosmological equation
of state as long as they can be associated with electromagnetic components.

The interpretation of the Big Rip as a one-way membrane analogous
to the infinite-redshift surface of a black hole is, as far as we
are aware, novel to this work. The continuity of the gravitational
wave function across the singularity might loosely be interpreted
as transmitting information to another universe which begins with
a Big Bang at $t=t_{rip}$, but the fact of the continuity of the
metric means we must be more precise about the meaning of ``universe''.
Without doubt, the Big Rip remains a catastrophic event for matter,
although transmission of gravitational waves generated by the motion
of that matter across the singularity might create a kind of immortality
in the form of persistent perturbations to the $t>t_{rip}$ region
of spacetime.

It is likely no coincidence that the main area where confluent Heun
functions have appeared as solutions in gravitational physics is in
the perturbations of the Kerr black hole \cite{Kerr Heun}. An interesting
project for further mathematical exploration might be to see if a
coordinate transformation will relate the Big Bang-Big Rip cosmology
to the Kerr metric; if so, cosmic acceleration might be described
as analogous to a cosmic angular momentum in a sense very different
from that of the ``rotating universe'' Bianchi VII models.

\section*{Appendix: approximating confluent Heun functions in the high-frequency
regime\label{sec:Appendix}}

The general theory of Heun functions remains largely incomplete \cite{Ronveaux}
and even the computational tools to investigate them numerically are
still being developed \cite{Fiziev}. Therefore for analytic work
it is necessary to make use of special values of the Heun functions'
parameters in order to relate the results to better-known cases.

To derive the approximations we make use of in this work, we are inspired
by the direction of \cite{Shahnazaryan-1}, who make use of transformations
of the confluent Heun equation in order to separate out one of the
equation's singular points. Consider the general form of the confluent
Heun equation:

\begin{equation}
f_{,z,z}+\left(4p+\frac{\gamma}{z}+\frac{\delta}{z-1}\right)f_{,z}+\frac{4p\alpha z-\sigma}{z\left(z-1\right)}f=0.\label{eq:canonical CHE}
\end{equation}
If we have $\gamma=\sigma=0$, or $\delta=4p\alpha-\sigma=0$, then
the singularity at $z=0$ or $z=1$ respectively vanishes and (\ref{eq:canonical CHE})
is reduced to a confluent hypergeometric equation. Our equation (\ref{eq:Heun wave})
and the confluent Heun equation related to (\ref{eq:rad heun t})
cannot be transformed by form-preserving transformations \cite{Maier}
into simpler equations, but they can be transformed into forms which
are \emph{almost} reducible, and which can be approximately solved.

Let $g\equiv\left(z-1\right)^{-\delta/2}f$. Then we have
\begin{equation}
g_{,z,z}+\left(4p+\frac{\gamma}{z}\right)g_{,z}+\frac{1}{z\left(z-1\right)^{2}}\left[\begin{array}{c}
4p\left(\alpha-\delta/2\right)\left(z-1\right)^{2}+\\
+\left(4p\alpha-2p\delta-\sigma-\gamma\delta/2+\delta/2-\delta^{2}/4\right)\left(z-1\right)+\\
+\delta/2-\delta^{2}/4
\end{array}\right]g=0.\label{eq:Heun separated 1}
\end{equation}
In the case of (\ref{eq:Heun wave}) for $w_{0}=1/3$ the fact that
$\alpha=0,\gamma=-\delta=2,\sigma=2\gamma p$ means (\ref{eq:Heun separated 1})
reduces to
\begin{equation}
g_{,z,z}+\left(4p+\frac{2}{z}\right)g_{,z}+\frac{4p\left(z-1\right)^{2}-2}{z\left(z-1\right)^{2}}g=0\label{eq:Heun separated 1-1}
\end{equation}
where \emph{p} is an arbitrary imaginary number. In any practical
case for gravitational wave astronomy, the quantity $\left\Vert 4p\left(z-1\right)^{2}\right\Vert \ggg2$
(recall $\omega\xi=n\eta$) so we can say 
\begin{equation}
zg_{,z,z}+\left(4pz+2\right)g_{,z}+4pg\approx0.\label{eq:Heun CHGE 1}
\end{equation}
Further simplifying matters, the solution to (\ref{eq:Heun CHGE 1})
is of the form $g\approx C_{1}\left(\phantom{}_{1}F_{1}\left(-a,-2a,w\right)\right)+C_{2}\psi\left(-a,-2a,w\right)$
which is a well-known identity for the Bessel functions, where $\phantom{}_{1}F_{1}\left(a,b,w\right)$
is Kummer's degenerate hypergeometric function and $\psi\left(a,b,w\right)$
is Tricomi's function \cite[chapter 13]{A&S} and \emph{a}~and \emph{b}~are
arbitrary parameters.

The necessary condition for the approximation we have just made use
of is that the \emph{p}-dependent terms cancel out everywhere except
the leading term in the polynomial forming the numerator of the third
term in (\ref{eq:Heun separated 1}); that is, assuming the parameters
$\alpha$, $\gamma$ and $\delta$ are independent of \emph{p}, we
need $4p\alpha-2p\delta-\sigma=0$ (in terms of Leaver's parameters
for the GSWE as used in \cite{Leaver} this is simply the condition
$B_{3}=0$). If we impose this condition then (\ref{eq:Heun separated 1})
leads to the equation
\begin{equation}
zg_{,z,z}+\left(4pz+\gamma\right)g_{,z}+4p\left(\alpha-\frac{\delta}{2}\right)g\approx0\label{eq:Heun HF approximation}
\end{equation}
 which has solution
\begin{equation}
g\approx C_{1}\left(\phantom{}_{1}F_{1}\left(\alpha-\frac{\delta}{2};\gamma;-4pz\right)\right)+C_{2}\psi\left(\alpha-\frac{\delta}{2};\gamma;-4pz\right).\label{eq:-22}
\end{equation}
For (\ref{eq:Heun wave}) we have $\alpha=0,\gamma=-\delta,p=\imath\omega/2,z=\xi$
which leads directly to the Bessel function solution (\ref{eq:Heun approximation}).

Recalling the asymptotic form for the Bessel functions for large \emph{w}
\cite[chapter 9]{A&S}
\begin{equation}
C_{1}J_{\kappa}\left(w\right)+C_{2}Y_{\kappa}\left(w\right)\approx\sqrt{\frac{\pi}{2w}}\left[C_{1}\sin\left(w-\frac{\pi}{4}-\frac{\pi}{2}\kappa\right)+C_{2}\cos\left(w-\frac{\pi}{4}-\frac{\pi}{2}\kappa\right)\right]\label{eq:-30}
\end{equation}
lets us then obtain (\ref{eq:Heun final eta approximation}).

In the case of (\ref{eq:rad heun t}) the approximation follows mostly
the same pattern as above. Applying the same transformation to the
confluent Heun equation and approximating in the limit of large $\upsilon$
we arrive at the confluent hypergeometric equation
\begin{equation}
xg_{,x,x}+\left(2\upsilon x+\frac{3}{2}\right)g_{,x}+\left(\upsilon^{2}+\frac{3}{2}\upsilon\right)g\approx0.\label{eq:-31}
\end{equation}
This equation can be solved in terms of the parabolic cylinder function
$U\left(a,z\right)$ as defined in \cite[chp 19]{A&S} and approximated,
using Darwin's approximation, as
\begin{equation}
g\approx e^{-\upsilon x}\frac{\left(1-x\right)^{-\frac{1}{4}}}{\sqrt{-4\upsilon x}}\left[\begin{array}{c}
c_{1}\cos\left(\frac{1}{4}\pi-\frac{1}{2}\pi\upsilon+\upsilon\sqrt{x\left(1-x\right)}+\upsilon\sin^{-1}\sqrt{x}\right)+\\
+c_{2}\cos\left(\frac{1}{4}\pi-\frac{1}{2}\pi\upsilon-\upsilon\sqrt{x\left(1-x\right)}-\upsilon\sin^{-1}\sqrt{x}\right)
\end{array}\right];\label{eq:-32}
\end{equation}
thus we arrive at (\ref{eq:rad heun approx}).

Equation (\ref{eq:delta x}) is more general than the GSWE and so
solutions cannot necessarily be expressed as Heun functions, but if
we assume$\beta$ is a rational number such that $\left(1+\beta\right)/\left(2+\beta\right)=u/v$
and $u,v\in\mathbb{Z}\backslash0$ we can transform it into an equation
with polynomial coefficients solvable by the Frobenius method with
the change of variables $x\rightarrow\tau^{v}/\left(\tau^{v}+1\right)$;
this maps the interval $x\subseteq\left(0,1\right)$ to the interval
$\tau\subseteq\left(0,\infty\right)$. While this transformation introduces
$v-1$ new singularities at $\tau=\sqrt[v]{-1}$ these are all at
non-real values of $\tau$ and, as the form of the transformed equation
shows, all interchangeable through Möbius transformations:
\begin{equation}
\begin{aligned}\tau\left(\tau^{v}+1\right)^{4}\nu_{,\tau,\tau}+\left(\tau^{v}+1\right)^{3}\left[\left(1+v+3u\right)\tau^{v}+\left(1-v+3u\right)\right]\nu_{,\tau}+\\
+v^{2}\upsilon^{2}\tau^{2v-2u-1}\nu & =0.
\end{aligned}
\label{eq:-20}
\end{equation}
This general form, while well-suited to numerical work, does not produce
intuitive analytic results so we turn again to approximation. If we
assume $0\leq\beta\leq1$, as we must for physically-interesting cases,
then to order $\tau^{v-1}$ we have a wave equation of the same form
as (\ref{eq:GW conformal equation}), leading to the approximate solution
(\ref{eq:General t solution}).

\section*{Acknowledgements}

The author would like to thank Dr Richard Schluessel \& Ms Cynthia
Conti for their support during the writing of this work.

Professor Plamen Fiziev of Sofia University provided helpful conversations
at a critical point in the exploration of this work's mathematics.

Numerical calculations were performed and plotted using Gnu Octave
3.6.4 with the signal-1.2.2 package.


\begin{thebibliography}{10}
\bibitem[1]{Riees acceleration}Riess, A et al. ``Observational Evidence
from Supernovae for an Accelerating Universe and a Cosmological Constant.''
The Astronomical Journal, Volume 116, Issue 3, pp. 1009-1038. (09/1998)
arXiv:astro-ph/9805201

\bibitem[2]{MTW}Misner, C \emph{et al.} (1973). Gravitation. W. H.
Freeman. p. 835

\bibitem[3]{Phantom energy}Caldwell, R. \emph{et al.} ``Phantom
Energy and Cosmic Doomsday''. Phys.Rev.Lett. 91 (2003) 071301. arXiv:astro-ph/0302506v1 

\bibitem[4]{Babichev}Babichev, E. \emph{et al.} ``Dark energy cosmology
with generalized linear equation of state''. Class.Quant.Grav.22:143-154,2005.
arXiv:astro-ph/0407190v3

\bibitem[5]{H0}Planck collaboration. ``Planck 2013 results. I. Overview
of products and scientific results''. arXiv:1303.5062v1

\bibitem[6]{q0}Giostri, R. \emph{et al. }``From cosmic deceleration
to acceleration: new constraints from SN Ia and BAO/CMB''. JCAP03(2012)027
arXiv:1203.3213v1

\bibitem[7]{eLisa}Amaro-Seaone, P \emph{et al.} ``Low-frequency
gravitational-wave science with eLISA/NGO''. Class. Quantum Grav.
29 124016 arXiv:1202.0839v2

\bibitem[8]{Slavyanov}Slavyanov, S \& Lay, W. \emph{Special Functions:
A Unified Theory Based on Singularities. }Chapter 3.1.2. Oxford University
Press, 2000.

\bibitem[9]{Leaver}Leaver, Edward W. \textquotedbl{}Solutions to
a generalized spheroidal wave equation: Teukolsky\textquoteright{}s
equations in general relativity, and the two-center problem in molecular
quantum mechanics.\textquotedbl{}\emph{ Journal of mathematical physics}
27 (1986): 1238.

\bibitem[10]{Probing spacetimes}Van Den Broeck, C. ``Probing dynamical
spacetimes with gravitational waves''. arXiv:1301.7291v1 {[}gr-qc{]}

\bibitem[11]{Confluent solutions}El-Jaick, Lea Jaccoud, and Bartolomeu
DB Figueiredo. \textquotedbl{}On Certain Solutions for Confluent and
Double-Confluent Heun Equations.\textquotedbl{} arXiv preprint arXiv:0807.2219
(2008).

\bibitem[12]{Shahnazaryan}Shahnazaryan, V \emph{et al.} ``On Certain
Solutions For Confluent And Double-confluent Heun Equations''.

\bibitem[13]{Fiziev}Fiziev, P. Novel relations and new properties
of confluent Heun\textquoteright{}s functions and their derivatives
of arbitrary order 

\bibitem[14]{GDI}Grishchuk, L. P., A. G. Doroshkevich, and V. M.
Iudin. \textquotedbl{}Long gravitational waves in a closed universe.\textquotedbl{}
Zhurnal Eksperimental'noi i Teoreticheskoi Fiziki 69 (1975): 1857-1871.

\bibitem[15]{Kerr Heun}Suzuki, H, Takasugi, E \& Umetsu, H. ``Perturbations
of Kerr-de Sitter Black Holes and Heun's Equations''

\bibitem[16]{Yang Zhang relic waves}Yang Zhang \emph{et al. }``Relic
gravitational waves in the accelerating universe''. arXiv:astro-ph/0501329v1

\bibitem[17]{Sako}Sako, M et al. ``The Sloan Digital Sky Survey-II
Supernova Survey: Search Algorithm and Follow-Up Observations''.
The Astronomical Journal, Volume 135, Issue 1, pp. 348-373 (2008).
(01/2008) arXiv:0708.2750

\bibitem[18]{Sollerman}Sollerman, J et al. ``First-Year Sloan Digital
Sky Survey-II (SDSS-II) Supernova Results: Constraints On Non-Standard
Cosmological Models''. The Astrophysical Journal, Volume 703, Issue
2, pp. 1374-1385 (2009). (10/2009) arXiv:0908.4276

\bibitem[19]{Mikinatis}Mikinatis, G et al. ``The ESSENCE Supernova
Survey: Survey Optimization, Observations, and Supernova Photometry''.
The Astrophysical Journal, Volume 666, Issue 2, pp. 674-693. (09/2007)
arXiv:astro-ph/0701043

\bibitem[20]{Sullivan}Sullivan, M et al.'' The dependence of Type
Ia Supernovae luminosities on their host galaxies''. Monthly Notices
of the Royal Astronomical Society, Volume 406, Issue 2, pp. 782-802.
(08/2010) arXiv:1003.5119

\bibitem[21]{Sullivan 2}Sullivan, M et al. ``SNLS3: Constraints
on Dark Energy Combining the Supernova Legacy Survey Three Year Data
with Other Probes''. The Astrophysical Journal, Volume 737, Issue
2, article id. 102 (2011). (08/2011) arXiv:1104.1444v2

\bibitem[22]{Blake}Blake, C et al. ``The WiggleZ Dark Energy Survey:
the growth rate of cosmic structure since redshift z = 0.9''. Monthly
Notices of the Royal Astronomical Society, Volume 415, Issue 3, pp.
2876-2891. (08/2011) arXiv:1104.2948v1

\bibitem[23]{Drinkwater}Drinkwater, M et al. ``The WiggleZ Dark
Energy Survey: Survey Design and First Data Release''. Monthly Notices
of the Royal Astronomical Society, Volume 401, Issue 3, pp. 1429-1452.
(01/2010) arXiv:0911.4246

\bibitem[24]{Perlmutter}Perlmutter, S et al. ``Measurements of Omega
and Lambda from 42 High-Redshift Supernovae''. Astrophysical Journal
517:565-586, 1999. (12/1998) arXiv:astro-ph/9812133v1

\bibitem[25]{Kowalski}Kowalski, M et al. ``Improved Cosmological
Constraints from New, Old, and Combined Supernova Data Sets.'' The
Astrophysical Journal, Volume 686, Issue 2, pp. 749-778. (10/2008)
arXiv:0804.4142v1

\bibitem[26]{Astier}Astier, P et al. ``The Supernova Legacy Survey:
measurement of $\Omega_{M}$, $\Omega_{\Lambda}$, and \emph{w} from
the first year data set. Astronomy and Astrophysics, Volume 447, Issue
1, February III 2006, pp.31-48. (02/2006) arXiv:astro-ph/0510447

\bibitem[27]{Wood-Vasey}Wood-Vasey, W et al. ``Observational Constraints
on the Nature of Dark Energy: First Cosmological Results from the
ESSENCE Supernova Survey.'' The Astrophysical Journal, Volume 666,
Issue 2, pp. 694-715. (09/2007) arXiv:astro-ph/0701041

\bibitem[28]{Davis}Davis, T et al. ``Scrutinizing Exotic Cosmological
Models Using ESSENCE Supernova Data Combined with Other Cosmological
Probes.'' The Astrophysical Journal, Volume 666, Issue 2, pp. 716-725.
(09/2007) arXiv:astro-ph/0701510

\bibitem[29]{Riess 2}Riess, A et al. ``Type Ia Supernova Discoveries
at from the Hubble Space Telescope: Evidence for Past Deceleration
and Constraints on Dark Energy Evolution.'' The Astrophysical Journal,
Volume 607, Issue 2, pp. 665-687. (06/2004) arXiv:astro-ph/0402512

\bibitem[30]{Riess 3}Riess, A et al. ``New Hubble Space Telescope
Discoveries of Type Ia Supernovae at : Narrowing Constraints on the
Early Behavior of Dark Energy.'' The Astrophysical Journal, Volume
659, Issue 1, pp. 98-121. (04/2007) arXiv:astro-ph/0611572

\bibitem[31]{Drinkwater 2}Drinkwater, M et al. ``The WiggleZ Dark
Energy Survey: survey design and first data release'' Monthly Notices
of the Royal Astronomical Society, Volume 401, Issue 3, pp. 1429-1452.
(01/2010) arXiv:0911.4246v2

\bibitem[32]{Blake 2}Blake, C et al. ``The WiggleZ Dark Energy Survey:
measuring the cosmic expansion history using the Alcock-Paczynski
test and distant supernovae'' (08/2011) arXiv:1108.2637v1

\bibitem[33]{Schluessel}Schluessel, E. ``Long-wavelength gravitational
waves and cosmic acceleration''. PhD Thesis, Cardiff University,
2011.

\bibitem[34]{Sachs-Wolfe}Sachs, R; Wolfe A. ``Perturbations of a
Cosmological Model''. Astrophysical Journal, vol. 147, p.73 (01/1967)

\bibitem[35]{Suzuki}Suzuki, N et al. ``The Hubble Space Telescope
Cluster Supernova Survey: V. Improving The Dark Energy Constraints
Above Z>1 And Building An Early-Type-Hosted Supernova Sample''. eprint
arXiv:1105.3470 (05/2011) arXiv:1105.3470v1

\bibitem[36]{LL}Landau, L; Lifshitz, E. \emph{The Classical Theory
of Fields, Fourth Revised English Edition}. Oxford: Pergamon Press
(1975)

\bibitem[37]{A&S}Abramowitz, M; Stegun, I, ed. \emph{Handbook of
Mathematical Functions}, ninth printing, National Bureau of Standards
Press (1970)

\bibitem[38]{Lifshitz}Lifshitz, EM. ``On the gravitational stability
of the expanding universe''. J. Phys USSR, 1946, vol\textbf{ }10,\textbf{
}116

\bibitem[39]{Ronveaux}Ronveaux, A \& Arscott, F. \emph{Heun's Differential
Equations. }Oxford University Press, 1995.

\bibitem[40]{Shahnazaryan-1}Shahnazaryan, V \emph{et al. ``}New
Relations For The Derivative Of The Confluent Heun Function''.\emph{
}Armenian Journal of Physics, 2012, vol. 5, issue 3, pp. 146-155 

\bibitem[41]{Maier}Maier, R. \textquotedbl{}The 192 solutions of
the Heun equation.\textquotedbl{} Mathematics of Computation 76.258
(2007): 811-843.

\bibitem[42]{Planck}Ade, P, \emph{et al.} ``Planck 2013 results.
XVI. Cosmological parameters.'' arXiv: 1303.5076. 2013.\end{thebibliography}
\end{document}